\documentclass{IEEEtran}
\pdfoutput=1
\usepackage{amsmath,amssymb,amsfonts}
\usepackage{algorithm}
\usepackage{algorithmicx}
\usepackage{algpseudocode}
\usepackage{graphicx}
\usepackage{textcomp}
\usepackage{subcaption}
\usepackage{multicol}
\usepackage{lipsum}
\usepackage{enumerate}
\usepackage{bm}
\usepackage{etoolbox}
\usepackage{array}
\usepackage[numbers]{natbib}
\newtheorem{lemma}{Lemma}
\newtheorem{theorem}{Theorem}

\newcounter{rowcntr}[table]
\renewcommand{\therowcntr}{\thetable.\arabic{rowcntr}}

\newcolumntype{N}{>{\refstepcounter{rowcntr}\therowcntr}c}

\AtBeginEnvironment{tabular}{\setcounter{rowcntr}{0}}

\newcommand{\RNum}[1]{\uppercase\expandafter{\romannumeral #1\relax}}

\def\BibTeX{{\rm B\kern-.05em{\sc i\kern-.025em b}\kern-.08em
    T\kern-.1667em\lower.7ex\hbox{E}\kern-.125emX}}
\begin{document}

\title{Expected Constant Time Self-stabilizing Byzantine Pulse Resynchronization
\thanks{This work has been submitted to the Elsevier for possible publication. Copyright may be transferred without notice, after which this version may no longer be accessible.}
}

\author{Shaolin Yu*, Jihong Zhu, Jiali Yang, Wei Lu\\Tsinghua University, Beijing, China \\ysl8088@163.com}

\maketitle

\begin{abstract}
In extending fast digital clock synchronization to the bounded-delay model, the expected constant time Byzantine pulse resynchronization problem is investigated.
In this problem, the synchronized state of the system should not only be deterministically maintained but be reached from arbitrary states with expected constant time in the presence of Byzantine faults.
An intuitive geometric representation of the problem is introduced, with which the classical approximate agreement, randomized Byzantine agreement, and random walk are integrated with some geometric operations.
Efficient realizations are also provided for practical uses.
Compared with the state-of-the-art solutions, the assumed common pulses need not be regularly generated, the message complexity can be lowered as approximate agreement, and the expected stabilization time is optimal.
With this, the provided solution can efficiently convert irregularly generated common pulses to self-stabilizing Byzantine pulse synchronization.
\end{abstract}

\begin{IEEEkeywords}
self-stabilization, pulse synchronization, approximate agreement, random walk, consensus
\end{IEEEkeywords}

\section{Introduction}
\label{sec:Introduction}
Distributed hard-real-time systems often rely on globally synchronized pulses or clocks to coordinate time-critical tasks performed in the distributed entities.
For example, TDMA communication systems are built upon globally synchronized communication rounds in achieving high bandwidth utilization and temporal fault isolation.
Synchronous fault-tolerant systems are built upon periodically generated sparse events and digital clock synchronization in running the synchronous distributed algorithms.
Some cyber-physical systems are built upon a global time reference in performing distributed measurements, controls, and various time-sensitive computation and communication tasks.
In practice, as these hard-real-time systems are often safety-critical, the fundamental problem is to provide both high-reliable and high-available synchronization services for various upper-layer dependable applications.

For high reliability, the synchronization systems are often required to be Byzantine-fault-tolerant.
Namely, the desired synchronization service should be provided in the presence of some nodes in the system being fail-arbitrarily (i.e., Byzantine).
To this end, optimal Byzantine resilience would be reached if $f$ Byzantine nodes can be tolerated in the system with $2f+1$ nodes being nonfaulty.
Meanwhile, for high availability, the synchronization systems are often required to be self-stabilizing.
Namely, with arbitrary initial states, all nonfaulty nodes should be globally synchronized in the desired stabilization time.
To this end, the stabilization time is expected to be as small as possible.
With these two fundamental requirements, self-stabilizing Byzantine-fault-tolerant clock synchronization (SSBCS) is expected to be reached as fast as possible with sufficiently high Byzantine resilience.

In handling the SSBCS problem, different system settings are assumed for providing various kinds of synchronization services.
For example, by assuming the existence of common pulses, digital SSBCS \citep{Hoch2006Digital,Ben2008Fast,DolevWelchSelf2004} aims to synchronously count the number of the globally generated common pulses in all nonfaulty nodes.
Under this basic setting, expected-constant-time optimal-resilient (\emph{ECTOR} in short) digital SSBCS can be reached in completely connected networks \citep{Ben2008Fast}.
Nevertheless, in building real-world digital SSBCS systems, the assumed common pulses should be well implemented with the underlying pulsing schemes.
Meanwhile, in constructing the desired real-time SSBCS system from the digital SSBCS and the common pulses, these pulses are often required to be periodically generated with adequate precision and accuracy.
Furthermore, for safety-critical applications, the underlying pulsing schemes are required to be reliable.
Thus, the fundamental problem is to provide periodically generated common pulses with adequate precision and accuracy in the distributed system with Byzantine-fault-resilience and fast stabilization.
This is referred to as the self-stabilizing Byzantine pulse synchronization (SSBPS) problem.
Now in handling the SSBPS problem, several solutions \citep{DolevWelchSelf2004,DaliotBiological2003,DolevPulseBoundedDelay2007,Dolev2014PulseGeneration,Lenzen2019AlmostConsensus} are provided without the aids of common pulses in the system.
However, unlike the digital SSBCS problem, no expected-constant-time SSBPS solution is provided in establishing the desired pulse synchronization in the presence of $f$ Byzantine nodes.
The state-of-the-art randomized SSBPS solution can reach an expected $O(\log f)$ stabilization time with optimal-resilience \citep{Lenzen2019AlmostConsensus}.
Nevertheless, this is still at the expense of a significant linear coefficient in the actual stabilization time and high message complexity.
Other linear-time SSBPS solutions, both the deterministic \citep{DolevPulseBoundedDelay2007} and the randomized \citep{Dolev2014PulseGeneration} ones, share the same disadvantage of the significant time coefficients.
Meanwhile, no sublinear-time deterministic SSBPS solution is known yet.
Thus, the questions posed in \cite{Ben2008Fast} \emph{if digital SSBCS can be transported to the bounded-delay model and reduce the convergence time to expected constant} and in \cite{Lenzen2019AlmostConsensus} \emph{if SSBPS is at least as hard as synchronous consensus} are still open.

In this paper, to partly answer these questions, instead of building the SSBCS and SSBPS systems from scratch, we explore how to relax the assumption of the common pulses while still making the SSBPS and SSBCS solutions being expected-constant--time (\emph{ECT} in short).
Concretely, we would still assume some common pulses are being globally generated.
Nevertheless, we allow these common pulses to be generated irregularly, as long as some relaxed liveness and separation conditions are satisfied.
Under this relaxed system setting, we would introduce the geometric representation of the self-stabilizing Byzantine pulse resynchronization (SSBPR) problem with geometric intuitions.
Then, by combining classical approximate agreement \citep{RN4093}, randomized Byzantine agreement \citep{FM1997}, random walk, and some simple geometric operations, we would show that \emph{ECT} SSBPR, \emph{ECT} SSBPS, and \emph{ECT} SSBCS can be reached without significant time coefficients and great message complexity, providing that the relaxed common pulses can be generated by some underlying pulsing schemes.

This work would benefit both the theoretical and practical aspects of SSBCS and the related studies.
Theoretically, in extending \emph{ECT} digital SSBCS to \emph{ECT} real-time SSBCS with the bounded-delay model, the regularly generated common pulses are relaxed to the irregularly generated ones.
Meanwhile, by integrating the geometric SSBPR with random walks, some new synchronization primitive is provided in reaching easier approximate agreements in circular spaces.
Practically, as the irregularly generated common pulses might be more easily implemented in real-world systems than the regularly generated ones with the same fault-assumption coverage \citep{Powell1992assumption}, the efficiency of the real-time SSBCS systems would be improved with the easier synchronization primitives.
Meanwhile, high synchronization precision can also be expected with some kinds of irregularly generated common pulses such as some redundant reference broadcasts \citep{elson2002fine}.

The rest of this paper is structured as follows.
The related work is presented in Section~\ref{sec:Related}.
The system model and the core problems are given in Section~\ref{sec:Model}.
In Section~\ref{sec:Resync}, the SSBPR problem is studied with geometric intuition and is solved with geometric embedding strategies.
Then, the analysis of the core algorithm is presented in Section~\ref{sec:Ana}.
Finally, we conclude the paper in Section~\ref{sec:Conclusion}.

\section{Related Work}
\label{sec:Related}
Self-stabilization (SS), Byzantine-fault-tolerance (BFT), and clock synchronization (CS) have been widely investigated in the distributed computing realm for more than 40 years.
However, the integration of SS, BFT, and CS is not straightforward.
In \cite{DolevWelchSelf2004}, the first two randomized SSBCS solutions are provided.
The one is with the assumption of common pulses, and the other is without it.
Both solutions are optimal-resilient, while both being with expected exponential stabilization time in playing the \emph{scheduler-luck games} \citep{Dolev1995luck} with the adversary.
Almost at the same time, the first deterministic SSBCS solution without the assumption of common pulses is provided in \cite{DaliotBiological2003}, while with the assumption of reliable broadcast.

In \cite{Ben2008Fast}, by assuming regularly generated common pulses, \emph{ECTOR} digital SSBCS is provided with employing \emph{ECTOR} randomized Byzantine agreement (RBA) as the core primitive.
Another classical randomized SSBPS solution is provided in \cite{Dolev2014PulseGeneration} without the assumption of common pulses but with an expected linear stabilization time.
One advantage of \cite{Dolev2014PulseGeneration} is that the Byzantine agreement (BA) is avoided, with which the required bandwidth and computation resources can be significantly reduced.
However, as far as we know, \emph{ECT} SSBCS is only reached with the assumption of regularly generated common pulses.
The state-of-the-art randomized SSBCS without common pulses reaches expected logarithmic stabilization time with a high linear coefficient on the asymptotical stabilization time \citep{Lenzen2019AlmostConsensus}.
For further reducing the stabilization time, no nontrivial lower bound is known for randomized SSBCS without regularly generated common pulses.

In the deterministic approaches, linear-time SSBCS solutions without common pulses are provided in \cite{Daliot2006Linear,DolevPulseBoundedDelay2007} with employing deterministic self-stabilizing BA (DSSBA) as the core primitive.
In \cite{Lenzen2019AlmostConsensus}, the author shows that the deterministic SSBCS problem is almost as easy as BA.
Unfortunately, as the message complexity and the required minimal rounds of BA are all lower-bounded with $f$, alternatives should be explored for deriving more efficient deterministic SSBCS solutions.
Also, no nontrivial lower bound of the stabilization time is known for deterministic SSBCS without common pulses.

In the industrial realm, most BFT CS solutions are not built upon full-BFT CS algorithms.
For example, the time-triggered architecture takes some light-weight Byzantine-fault-tolerant startup procedures \citep{Steiner2008StartupRecovery} in which some local central guardians or hardware monitor-pairs \citep{as6802} are employed.
Some other self-stabilizing CS solutions such as \cite{YuCOTS2021} can only tolerate the Byzantine faults generated in some specified nodes.

\section{System Model and Core Problems}
\label{sec:Model}
\subsection{The bounded-delay model}
The bounded-delay model is a good abstraction of real-world message-passing systems built upon indeterministic-delayed communication, limited-power computation, imperfect clocks, and unreliable components.
Meanwhile, this model provides basic system settings for building self-stabilizing Byzantine-fault-tolerant systems.
Following the SSBPS solutions provided in \cite{Daliot2006Linear,DolevPulseBoundedDelay2007,Dolev2014PulseGeneration,Lenzen2019AlmostConsensus}, this paper also takes the bounded-delay model in handling the core problems.

In the core abstraction, the completely-connected communication network $G=(N,N\times N)$ of the system $\mathcal{S}$ consists of $n=|N|$ nodes.
Faulty and nonfaulty nodes in $N$ are denoted as $F$ and $Q=N\setminus F$, respectively.
The physical time (real-time) and readings (of ticks) from the local clock of node $q\in N$ (local-time of $q$) are respectively denoted as $t$ and $\tau_{q}$.
A node $q\in N$ is nonfaulty if and only if (iff) $q$ processes and sends messages according to the provided algorithms, the drift-rate of $\tau_{q}$ (between adjacent overflows) is bounded within $[-\rho,\rho]$ with respect to $t$, and the overall processing delay of a message in $q$ is less than $d_\mathtt{p}$ real-time units.
A faulty node can fail arbitrarily and send arbitrary messages or nothing to any subset of $N$ at any time under the full control of a static adversary (i.e., the adversary cannot change the faulty nodes during any execution of $\mathcal{S}$ but can arbitrarily choose $f$ nodes in $N$ being faulty at the beginning of each such execution).
The communication network is nonfaulty at $t_0$ iff all messages sent at or before $t_0$ can arrive their destination nodes before $t_0+d_\mathtt{m}$.
Denoting $d=d_\mathtt{m}+2d_\mathtt{p}$, $\mathcal{S}$ is said to be nonfaulty \emph{since} $t_0+d$ iff the communication network is nonfaulty and $|F|\leqslant f$ \emph{since} $t_0$ (a condition holds \emph{since} $t_0$ iff it holds at all $t\in[t_0,t_{curr}]$ where $t_{curr}$ is the current time).
In all cases discussed in the paper, we assume $f<n/3$, i.e., the optimal Byzantine resilience.
In some specific cases, we relax this condition to $f=O(\sqrt{n})$.

Inevitably, the local-time $\tau_{q}$ of a node $q\in Q$ has an upper-bound $\tau_{max}-1$ after which $\tau_{q}$ would return to $0$, where $\tau_{max}$ is assumed large enough to run the timers in the provided algorithms.
Meanwhile, in the words of self-stabilization, $\mathcal{S}$ is not always nonfaulty as there could be transient system failures during which the communication network could fail arbitrarily, and $|F|$ could be up to $n$.
Thus, when $\mathcal{S}$ becomes nonfaulty at $t_0$, $\tau_{q}$ and other local variables (not including the constant parameters) could take arbitrary values in their valid ranges.
In the overall SSBPS system, as we consider the adversary as static, when $\mathcal{S}$ is nonfaulty, the set of the faulty nodes $F$ would remain to be unchanged.
For simplicity and without loss of generality, we use $Q=\{1,2,\dots,|Q|\}$ and $F=\{|Q|+1,\dots,n\}$ to respectively denote the nonfaulty and faulty nodes in the nonfaulty $\mathcal{S}$.
Meanwhile, $\mathcal{S}$ is assumed to be nonfaulty since $t=0$.

\subsection{The pulsing systems}
Generally, the so-called \emph{pulse} refers to some specific event that is instantly generated in a nonfaulty node at some \emph{pulsing instant}.
Denoting the $k$th ($k\geqslant 0$) pulsing instant of node $q$ since $t_0$ as $t_{q}^{(t_0,k)}$, $\mathcal{S}$ is a $(\Pi, T, \Delta)$-pulsing system since $t$ iff for all $q,q'\in Q$ and $k\geqslant 0$
\begin{eqnarray}
\label{eq_sync_precision}
1)~\textit{Precision:}~ |t_{q}^{(t,k)}-t_{q'}^{(t,k)}|\leqslant \Pi ~~~~~~~~~~~\\
\label{eq_sync_accuracy}
2)~\textit{Accuracy:}~ T-\Delta \leqslant t_{q}^{(t,k+1)}-t_{q}^{(t,k)}\leqslant  T+\Delta
\end{eqnarray}
hold, where the finite parameters $\Pi$, $T$, and $\Delta$ are respectively the synchronisation precision, the nominal pulsing cycle, and the maximal stable pulsing jitter measured in real-time.
To exclude trivial solutions, $\Pi\ll T-\Delta$ is often required.
Furthermore, to provide regularly generated common pulses in SSBPS, $\Pi$ and $\Delta$ are often required to be as small as possible.
In this paper, the desired $(\Pi, T, \Delta)$-pulsing system is denoted as $\mathcal{S}_1$.

In building $\mathcal{S}_1$ with a small $\Delta$, we assume that a relaxed $(\Pi_0, T_0, \Delta_0)$-pulsing system $\mathcal{S}_0$ with a much larger $\Delta_0$ has been built in $\mathcal{S}$ since $t=0$.
For $T_0$ and $\Delta_0$, we only require $T_0-\Delta_0>\Gamma_0$, where $\Gamma_0$ is the minimal time for running some pulse resynchronization routine.
With this, $T_0$ can be independent of $T$ and $\Delta_0$ can be arbitrarily large, as long as $T_0-\Delta_0>\Gamma_0$ and the accumulated drifts are allowed in $\Pi$.
Meanwhile, $\Pi_0$ is still required to be as small as possible.
With this, the pulses generated in $\mathcal{S}_0$ are called the irregularly generated common pulses (\emph{igc} pulses).

To represent the pulsing state in building $\mathcal{S}_1$, we use
\begin{eqnarray}
\label{eq_localphase}
\theta_q(t)=(T-((\tau_{sch}^{(q)}(t)-\tau_q(t))\bmod\tau_{max}))\bmod T
\end{eqnarray}
to represent the \emph{pulsing phase} of $q$ at real-time $t$, where $\tau_{sch}^{(q)}(t)$ is the scheduled pulsing tick in $q$ at $t$ for generating the next pulse.
Namely, when $\tau_q=\tau_{sch}^{(q)}$, $q$ would generate a pulse and reset $\tau_{sch}^{(q)}$ as $(\tau_{sch}^{(q)}+T)\bmod \tau_{max}$.
For simplicity, we assume that every $q\in Q$ has one and only one such scheduled pulsing tick at any given instant $t\geqslant 0$.
By assuming $\tau_q$ and $\tau_{sch}^{(q)}$ are atomically updated, $\theta_q(t)$ is uniquely defined at any $t\geqslant 0$.

With this, the synchronized states of the pulsing system are the ones in which the pulsing phases of all nonfaulty nodes are sufficiently near at the same time.
In measuring how close two pulsing phases are, the value-space of the pulsing phases can be represented as a circular space.
For example, the distance between the pulsing phases $0$ and $ T-\epsilon$ should be $\epsilon$ rather than $ T-\epsilon$ for $0<\epsilon\ll  T$.
Without loss of generality, we can represent the pulsing phase $\theta$ as a point on a circle $C$ with unified circumference $|C|=1$.
By setting the point of $\theta=0$ as the origin (on somewhere of $C$), every point in $C$ can be represented by a number $c \in [0,1)$ that is the length of the shortest \emph{counterclockwise arc} (\emph{cc-arc} for short) from the origin to the point.
However, as $C$ is not homeomorphic to $[0,1)\subset \mathbb R$, the distance defined on $[0,1)$ cannot be directly applied on $C$.
In the former example, at any instant $t$, when the pulsing phases $\theta_{q_1}(t)$ and $\theta_{q_2}(t)$ of two nodes $q_1,q_2\in Q$ are respectively $0$ and $T-\epsilon$ with $\epsilon$ being a very small positive number, $\theta_{q_1}(t)$ and $\theta_{q_2}(t)$ should be regarded as being very near to each other, since it says that $q_1$ and $q_2$ are scheduled to generate their pulses in the desired synchronous way.
Also, there is no reason why $C$ should break at the point of $\theta=0$ when applying the FTA functions.
Namely, any point on $C$ might be chosen as the specific point to topologically break the circle $C$ into a segment for correctly applying the FTA functions, especially when $\mathcal{S}$ is synchronized.
Thus, the intuition is that the pulsing phases should be topologically represented on a circle.

So, to represent a cc-arc $\iota$ ending with two points $c_0,c_1\in C$, we slightly abuse the interval $[c_0,c_1]$ defined on real numbers by allowing both $c_0\leqslant c_1$ and $c_0>c_1$.
Namely, $\iota=[c_0,c_1]$ represents the shortest cc-arc on $C$ from point $c_0$ to $c_1$.
In particular, $[c,c]=\{c\}$ can also be represented by the single point $c$.
We say $\iota_1\subseteq \iota_2$ iff $\forall c\in \iota_1: c\in \iota_2$.
By defining $c_0\oplus c_1=(c_0+c_1)\bmod |C|$ and $c_0\ominus c_1=(c_0-c_1)\bmod |C|$, the length of a cc-arc $\iota=[c_0,c_1]$ can be represented as $|\iota|=c_1\ominus c_0$, the distance between two points $c_0,c_1\in C$ can be defined as $\mathring{d}(c_0,c_1)=\min\{c_1\ominus c_0,c_0\ominus c_1\}$, and the add operation on two cc-arcs can be defined as $\iota_1 + \iota_2 = \{c_1\oplus c_2\mid c_1\in \iota_1 \land c_2\in \iota_2\}$.
It is easy to see the function $\mathring{d}$ satisfies the criterion of distance, and the add operation of cc-arcs is commutable.
With this, the $\delta$-neighbourhood of a point $c\in C$ can be defined as a cc-arc $\kappa(c,\delta)=\{c'\mid \mathring{d}(c,c')\leqslant \delta\}$.
For any $S\subseteq C$, a cc-arc $\iota$ is a cover arc of $S$ iff $S\subseteq \iota$.
The length of the shortest cover arc of $S$ is denoted as $l(S)$.
It is easy to see that when $l(S)< 0.5$, the shortest cover arc of $S$ is unique and can be denoted as $\iota(S)$.
And for completeness, we set $\iota(S)=C$ when $l(S)\geqslant 0.5$.
Denoting the points corresponding to the pulsing phases of all nodes in $Q$ at real-time $t$ as $S_Q(t)=\{\theta_q(t)/ T\mid q\in Q\}$, we simply denote $\iota(S_Q(t))$ as $\iota_Q(t)$.
As the value-space of the pulsing phases is homeomorphic to $C$, it shares the definitions on $C$ by replacing $C$ as $[[T]]$, where $[[i]]=\{0,1,\dots,i-1\}$ is the set of the first $i$ natural numbers.

\subsection{The core problems}
\label{subsec:resynchronization}

Pulse synchronization is trivial in always-running synchronous systems whose synchronous round cycles can be directly utilized as the desired pulsing cycles.
Even with larger (or smaller) pulsing cycles being some multiples (or factors) of the synchronous round cycles, this problem can also be solved in expected constant time with fast digital clock synchronization (and the clocks with bounded drift-rates) \citep{Ben2008Fast}.
Furthermore, real-time SSBCS can also be built upon digital SSBCS with desired pulsing cycles.
However, it is not easy to establish such a synchronous system with the desired synchronous round cycles before reaching pulse synchronization.
Alternatively, it might be easier to generate some well-separated resynchronization pulses, with which a finite number of temporally established semi-synchronous rounds may follow to perform some pulse resynchronization routine.
For example, in \cite{Daliot2006Agreement}, DSSBA is built upon a finite number of semi-synchronous rounds of BA established with the $\mathtt{Initiator}$-$\mathtt{Accept}$ primitive.
In this case, the acceptance of the $\mathtt{Initiator}$-$\mathtt{Accept}$ primitive can be viewed as some resynchronization pulses.
However, as the $\mathtt{Initiator}$-$\mathtt{Accept}$ primitive cannot always be accepted with the desired initiation time, it does not support \emph{igc} pulses and would depend on some specific properties of the simulated synchronous BA routines \citep{Toueg1987Fast} in reaching the final agreement.
To avoid employing BA in the resynchronization routines, we assume the existence of $\mathcal{S}_0$.
With this, the SSBPS problem can be divided into two subproblems.
The one is to realize $\mathcal{S}_0$.
The other one is to realize $\mathcal{S}_1$ with $\mathcal{S}_0$.

Then, in realizing $\mathcal{S}_0$, the traditional theoretical problem is to provide the corresponding SSBPS algorithms for the unreliable computation nodes.
However, although this approach can often guarantee high assumption coverage, it is hard (if not impossible) to reach faster stabilization than the state-of-the-art solutions.
Alternatively, to fast generate \emph{igc} pulses with a sufficiently high assumption coverage, some simple realizations like the redundant reference broadcasts \citep{elson2002fine} may be the more practical options.
As the \emph{igc} pulses are often easier to be generated than the common pulses with high accuracy, the assumption coverage of the redundant pulsing schemes can be expected to be improved.
Then, by assuming that $\mathcal{S}_0$ is realized in some way, the remaining problem is to realize $\mathcal{S}_1$ with $\mathcal{S}_0$.

\section{Semi-Synchronous Pulse Resynchronization}
\label{sec:Resync}
To realize $\mathcal{S}_1$ with $\mathcal{S}_0$, the synchronized state of $\mathcal{S}_1$ should be reached and maintained with the \emph{igc} pulses.
For this, an \emph{igc} pulse is generated for every nonfaulty node $q\in Q$ at the beginning of every resynchronization round (r-round) of $q$.
Then, by counting the ticks passed in $q$, the $k_s$th ($1\leqslant k_s\leqslant K_{s}$) semi-synchronous round (ss-round) in this r-round of $q$ begins when $(k_s-1)\Phi_{s}$ ticks have passed since the beginning of the r-round.
Denoting $t_{q}^{(k,k_s)}$ as the instant at which the $k_s$th ss-round in the $k$th r-round begins, the ss-rounds are globally synchronized in that $\forall q',q\in Q, 1\leqslant k_s\leqslant K_{s}+1: |t_{q'}^{(k,k_s)}-t_{q}^{(k,k_s)}|\leqslant \pi_s$ and $\forall q\in Q, 1\leqslant k_s\leqslant K_{s}:\Phi_{s}^{-}\leqslant t_{q}^{(k,k_s+1)}-t_{q}^{(k,k_s)}\leqslant\Phi_{s}^{+}$.
During each ss-round, $q$ can exchange messages with other nodes and finish all related operations within $d$.
To separate messages exchanged in adjacent ss-rounds, $\Phi_{s}^{-}$ is assumed larger than $2\pi_s+d$.
For convenience, the ss-rounds in an r-round are also called the steps.
With $K_s \Phi_{s}^{+}<\Gamma_0$ and $\pi_s= \Pi_0+2\rho\Gamma_0$, a $K_s$-step resynchronization routine $\mathtt{Resync}$ can run in every r-round.

Then, denoting the end instant of the $k$th r-round in $q$ as $t_{q}^{(k)}$, each node $q\in Q$ adjusts its pulsing phase $\theta_{q}(t_{q}^{(k)+})=\theta_{q}(t_{q}^{(k)-})\oplus\Delta \theta_{q}^{(k)}$ by adjusting $\tau_{sch}^{(q)}$ as $(\tau_q+ ((-\theta_{q}(t_{q}^{(k)+}))\bmod T))\bmod \tau_{max}$, where $t^{+}$ and $t^{-}$ respectively denote the instant after and before $t$ with a infinitesimal duration, and $\Delta \theta_{q}^{(k)}$ is the output of $\mathtt{Resync}$.
Denoting $t_{first}^{(k)}=\min_{q\in Q}t_{q}^{(k)}$ and $t_{last}^{(k)}=\max_{q\in Q}t_{q}^{(k)}$, the interval $A_k=[t_{first}^{(k)},t_{last}^{(k)}]$ is referred to as the adjusting span in r-round $k$.
For convenience, we denote $t_{last}^{(0)}=t_{first}^{(1)}$.
Now in the $k$th r-round with $k\geqslant 1$, it is expected that
\begin{eqnarray}
\label{eq_Convergency}
1)~{Convergency:}~|\iota_Q(t_{last}^{(k)+})|\leqslant \epsilon_1\\
\label{eq_Continuity}
2)~{Continuity:}~|\iota_Q(t_{last}^{(k-1)+})|\leqslant \epsilon_1 \to \nonumber\\
\iota_Q(t_{last}^{(k)+})\subseteq \iota_Q(t_{first}^{(k)-})+\kappa(a_k,\rho\delta_s+\epsilon_2 )
\end{eqnarray}
where the unified adjusting span $a_k=|A_k|/ T$ is bounded by $\delta_s=\pi_s/ T$, and $\epsilon_1$, $\epsilon_2$ are respectively the upper-bounds of the unified resynchronization errors and accuracy errors.

In these two desired conditions, as the synchronized state of $\mathcal{S}_1$ should be deterministically maintained, (\ref{eq_Continuity}) should be satisfied in every r-round.
Nevertheless, as we want to avoid deterministic BA (DBA), we do not require (\ref{eq_Convergency}) to be satisfied in all cases.
Instead, when the system is not synchronized, several adjacent r-rounds can be viewed as the multi-round resynchronization progress, in which (\ref{eq_Convergency}) is expected to be satisfied before the last r-round.
Nevertheless, when the system is synchronized, (\ref{eq_Convergency}) should be satisfied deterministically.
Thus, the core problem is to design the $K_s$-step resynchronization routine $\mathtt{Resync}$ running in every r-round.
This is referred to as the SSBPR problem.

\subsection{Geometry of the resynchronization progress}
\begin{figure}[htbp]
\centering
\begin{subfigure}{.49\textwidth}
\centering\includegraphics[width=1.5in]{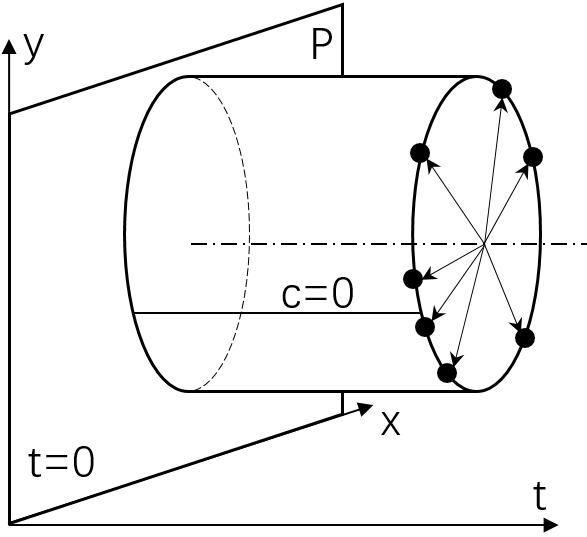}
\caption{The cylinder $C\times \mathbb{R}$ in $\mathbb{R}^3$.}
\label{fig:strategy_r3}
\end{subfigure}
\begin{subfigure}{.49\textwidth}
\centering\includegraphics[width=2.7in]{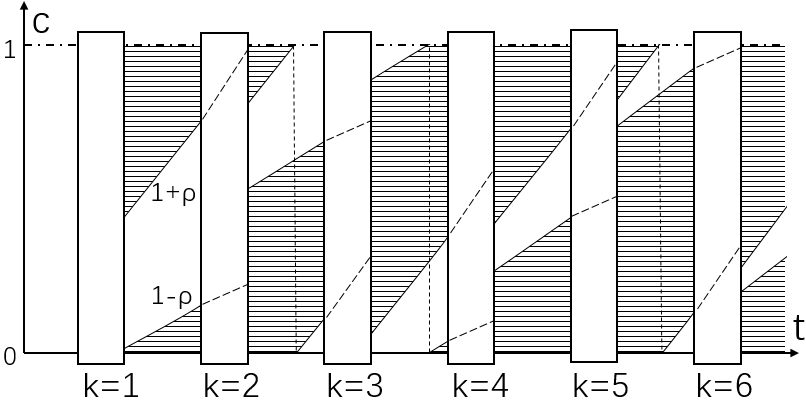}
\caption{The SSBPR progress on the cylinder $C\times \mathbb{R}$}
\label{fig:strategy_resync}
\end{subfigure}
\caption{Geometric representations of the problem}
\label{fig:regions}
\end{figure}

Intuitively, as the value-space of the pulsing phases is homeomorphic to the circle $C$, the SSBPR progress can be represented on a cylinder $C\times \mathbb{R}$ (the product of the unified circle $C$ and the real-time $t$) in $\mathbb{R}^3$, as is shown in Fig.~\ref{fig:strategy_r3}.
This cylinder is cut open along the line $c=0$ and thus can be flattened in Fig.~\ref{fig:strategy_resync}.

In Fig.~\ref{fig:strategy_resync}, each adjusting span is represented as a rectangular (referred to as the $k$th lens) with a unified length $\delta_s$.
The light (unshaded) areas are made of beams of light on $C\times \mathbb{R}$ which correspond to all possible pulsing phases of nodes in $Q$ along real-time $t$.
The shaded areas represent the impossible pulsing phases of the nonfaulty nodes.
As we assume that the drift rates of the local clocks are bounded within $[-\rho,\rho]$, the boundaries of the light areas and the shaded areas between the lenses are represented as lines.
In the progress of real-time $t$, the light beams between any two adjacent lenses would grow wider and wider with the bounded clock drifting effect.
So, it is desired that the light beams can become narrower at the right surface of each lens for reaching \emph{convergency}.
For \emph{continuity}, as a synchronized system should both maintain the \emph{Precision} and \emph{Accuracy} properties required in (\ref{eq_sync_precision}) and (\ref{eq_sync_accuracy}), the pulsing phases of all nodes in $Q$ should not be changed abruptly during an adjusting span.
So, once the light beam reaches the left surface of a lens with a facula no wider than $\epsilon_1$, it should leave the right surface of the lens within a facula being \emph{refracted} from the left surface.
Note that in Fig.~\ref{fig:strategy_resync}, the distances between the adjacent lenses are nearly identical.
This is not a necessity, as we only require \emph{igc} pulses.

Now, in providing the desired resynchronization while avoiding BA as much as possible, a basic strategy is to exchange pulsing phases and execute a fault-tolerant approximate (FTA) function to reach approximate agreement \citep{RN4093} of pulsing phases at the end of the $K_{s}$th ss-round.
However, as the pulsing phase $\theta_{q}$ of every node $q\in Q$ is continuously changing and is discontinuously valued around the maximal and minimal phases, the approximate function should work correctly in all possible cases.
For this, classical FTA functions would not do because there could be jumping changes of the averages when all pulsing phases of the nonfaulty nodes are around the maximal and minimal values (such as the cases of $k=3$ and $k=6$ in Fig.~\ref{fig:strategy_resync}).
However, if without the discontinuous presentation of the values, as all points of the initial pulsing phases could be symmetrically arranged on the circular space $C$ with distance $\mathring{d}$, the approximate function also needs some asymmetric operation in breaking this symmetry.

\subsection{Basic resynchronization strategies}
In tackling this dilemma with the resynchronization routine $\mathtt{Resync}$, we represent the cylinder $C\times \mathbb{R}$ in $\mathbb R^3$, as is shown in Fig.~\ref{fig:strategy_r3}.
With this, an intuitive conception for reaching SSBPR is space-embedding.
Namely, at any instant $t\geqslant 0$, in avoiding to directly handle the resynchronization problem on the circular space $C$, we can try first to map $C^{|Q|}$ to $P^{|Q|}$, then to apply some FTA functions on $P^{|Q|}$ to derive some $\mathbf{P}_{\varepsilon}=\{\vec{p}\in P^{|Q|} \mid r\in \mathbb{R}\land \|\vec{p}-r\cdot\vec{1}\|_{\infty}\leqslant \varepsilon/2\}$, and finally to map $\mathbf{P}_{\varepsilon}$ to $\mathbf{C}_{\epsilon_1}$, where $P$ is the required convergence space in applying the FTA functions.

\begin{figure}[htbp]
\centerline{\includegraphics[width=3.5in]{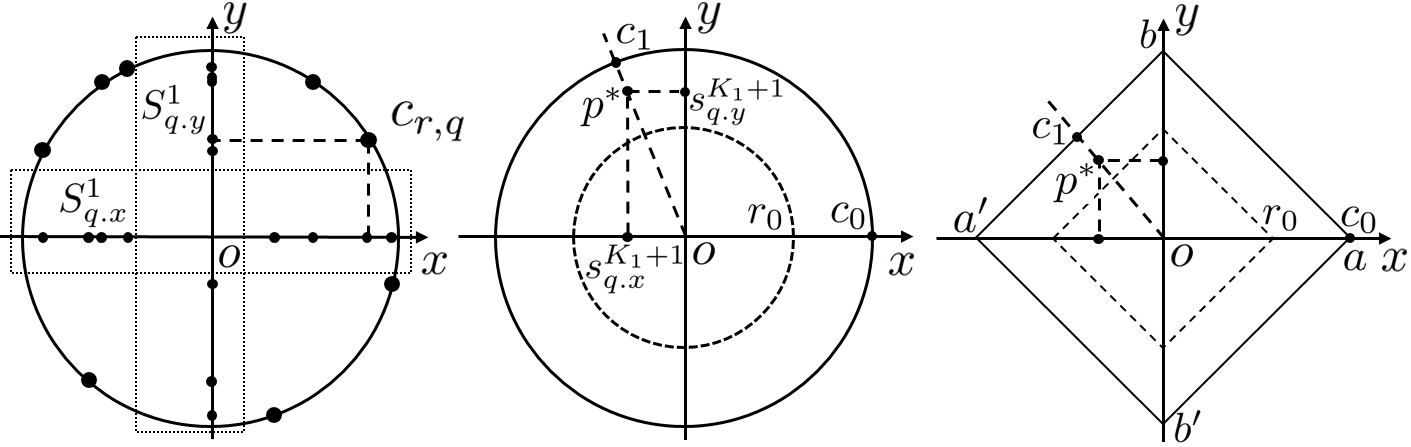}}
\caption{The embedding strategies on $\mathbb R^2$.}
\label{fig:ftcc0}
\end{figure}

Concretely, the $\mathtt{Resync}$ routine running in each r-round of the SSBPR progress can be constructed as follows.
Firstly, the $\mathtt{Resync}$ routine is composed of two successive stages with respectively $K_1$ and $K_2=K_s-K_1$ steps.
During the first stage, each node $q\in Q$ exchanges its current pulsing phase $\theta_{q}$ with all other nodes at the first step.
Denoting the message received from node $r\in N$ in node $q\in Q$ at this step as $\theta_{r,q}$, we know that $\theta_{r,q}$ would be valued as $\theta_{r}$ for $r\in Q$ and would be arbitrarily valued in $[[T]]$ for $r\notin Q$.
In case no correct message is received from node $r$ at the end of a step, $q$ can use its own value as the default value of $r$ (such as setting $\theta_{r,q}=\theta_{q}$) in this step.
Then, by denoting $c_{r,q}=\theta_{r,q}/ T$ as the unified pulsing phase of $r$ in $q$, the messages received from all distinct $n$ nodes in node $q\in Q$ at the first step can be unified as a point set $S_q=\{c_{r,q}\mid r\in N\}$, as is shown in Fig.~\ref{fig:ftcc0}.

In the leftmost of Fig.~\ref{fig:ftcc0}, projections of $S_q$ to the $ox$ and $oy$ axes on the $oxy$ plane $P$ ($P$ is also shown in Fig~\ref{fig:strategy_r3}) are shown as $S_{q.x}^{1}$ and $S_{q.y}^{1}$ respectively.
Then, some FTA functions are executed on these projections rather than on the points on $C$.
Namely, the projections $S_{q.x}^{1}$ and $S_{q.y}^{1}$ on the $ox$ and $oy$ axes are respectively used to compute the next round value $(s_{q.x}^{2},s_{q.y}^{2})=(\mathtt{FTA}(S_{q.x}^{1}),\mathtt{FTA}(S_{q.y}^{1}))$ with an FTA function $\mathtt{FTA}$.
Iteratively, at the $k$th ($1<k\leqslant K_1$) step, each node $q$ exchanges $(s_{q.x}^{k},s_{q.y}^{k})$ with all other nodes and gets $S_{q.x}^{k}=\{s_{r.x}^{k}\mid r\in N\}$ and $S_{q.y}^{k}=\{s_{r.y}^{k}\mid r\in N\}$.
With this, the next round value $(s_{q.x}^{k+1},s_{q.y}^{k+1})$ is computed as $(\mathtt{FTA}(S_{q.x}^{k}),\mathtt{FTA}(S_{q.y}^{k}))$ in $q$.
Then, at the end of the $K_1$th step of the first stage, the new computed round value $(s_{q.x}^{K_1+1},s_{q.y}^{K_1+1})$ in each node $q\in Q$ can be viewed as an \emph{anchor point} $p^*=(x^*_q,y^*_q)$ on the convergence plane $P$.

In the middle of Fig.~\ref{fig:ftcc0}, it is easy to see that if $|op^*|\neq 0$, the line $op^*$ and the circle $C$ has one and only one intersection point $c_\mathtt{1}$ on $P$ in node $q$.
In this case, the intersection point $c_\mathtt{1}$ is referred to as the ashore point on $C$.
With this, if all anchor points of nodes in $Q$ can be confined in an $\epsilon\times\epsilon$ square with the sides parallel to the $ox$ and $oy$ axes on $P$ (referred to as an $\epsilon$-square), the corresponding ashore points would be within an arc of length approximately proportional to $\epsilon/|op^*|$, providing that $|op^*|$ is sufficiently larger than $\epsilon$.
However, if $|op^*|=0$ (or being relatively small), the ashore points of all nodes in $Q$ cannot be covered in a short arc on $C$.
In this case, the ashore point $c_\mathtt{1}$ cannot be used as the reference point.

During the second stage of the r-round, denoting the significance of $p^*$ as $c_\mathtt{w}\in [0,1]$, every $q\in Q$ would set $c_\mathtt{w}$ according to the position of $p^*$.
Then, the reference point $c^*$ would be computed as
\begin{eqnarray}
\label{eq_creference}
c^*=c_\mathtt{w} c_\mathtt{1}
\end{eqnarray}

Firstly, with some threshold $r_0>0$, we might want to set $c_\mathtt{w}=1$ if $|op^*|>r_0$ and set $c_\mathtt{w}=0$ otherwise.
However, as we also want to avoid DBA, the nodes in $Q$ cannot always reach an exact agreement on the condition $|op^*|>r_0$.
In overcoming this, a straightforward idea might be to run a $K_2$-step protocol $\mathtt{A}$ in $Q$ to decide if the ashore point $c_\mathtt{1}$ can be used as the reference point.
Concretely, the common-coin-based constant-round randomized protocol $P_r$ provided in \cite{FM1997} (not the eventual RBA protocol) can be employed here as the desired $\mathtt{A}$ for reaching an agreement of the significance of $p^*$ with a probability no less than $0.35$, as the oblivious common coin is $0.35$-fair.
Then, at the end of the second stage of the r-round, $q$ can set $c^*=c_\mathtt{1}$ when $\mathtt{A}$ outputs $1$ and set $c^*=c_\mathtt{0}=0$ otherwise, where the origin point $c_\mathtt{0}$ can be any fixed point on $C$.
For example, in Fig.~\ref{fig:ftcc0}, the origin point $c_\mathtt{0}$ is chosen as the intersection point of $C$ and the $ox$ axis.
In this way, $c^*$ can be used as the reference point in $q$ at the end of the r-round.
As the first and second stages are all with constant rounds, $K_s=K_1+K_2$ is independent of $f$, and the solution can be optimal Byzantine resilient.
However, the randomized protocol $P_r$ provided in \cite{FM1997} (and the former one in \cite{FM1989} with $f<n/4$) requires high message complexity.

Alternatively, as the condition $|op^*|>r_0$ cannot always be consistently determined in all nodes of $Q$ without reaching DBA, we want to employ some composed thresholds rather than just $r_0$.
Concretely, by defining the significance $c_\mathtt{w}$ as
\begin{eqnarray}
\label{eq_c_w}
c_\mathtt{w}=\left\{
\begin{aligned}
1 &, & {|op^*|\geqslant r_0'}\\
b_\mathtt{w} &, & {r_0<|op^*|<r_0'}\\
0 &, & {|op^*|\leqslant r_0}
\end{aligned}
\right.
\end{eqnarray}
with some $r_0'>r_0>0$, the core problem is to determine $b_\mathtt{w}$ when $r_0<|op^*|<r_0'$.

In the simplest way, to determine such $b_\mathtt{w}$, each node $q\in Q$ can first randomly toss an unbiased coin $\bar{b}_q\in \{-1,1\}$.
Then, $q$ can exchange $\bar{b}_q$ with all the other nodes and collect a boolean vector $\tilde{B}=(\tilde{b}_1,\tilde{b}_2,\dots,\tilde{b}_n)$, where the collected bit $\tilde{b}_i$ would be $\bar{b}_i$ if $i\in Q$ and would be arbitrarily valued in $\{-1,1\}$ if $i\in F$.
Then, with $f=O(\sqrt{n})$, $q$ can simply compute $b_\mathtt{w}$ as
\begin{eqnarray}
\label{eq_b_w}
b_\mathtt{w}=\left\{
\begin{aligned}
1 &, & \sum_{i\in N}\tilde{b}_i>0\\
0 &, & \sum_{i\in N}\tilde{b}_i\leqslant 0
\end{aligned}
\right.
\end{eqnarray}

With this basic random walk strategy, the second stage of an r-round needs only $K_2=1$ step (for generating $b_\mathtt{w}$).
In the next section, we would show that such generated $b_\mathtt{w}$ can make the reference point $c^*$ of all nonfaulty nodes being covered in a sufficiently short arc on $C$ with some positive probability.
Here it should be noted that the coins should be tossed after the end of the first stage in every r-round.
\subsection{Easy realizations}
\begin{figure}[htbp]
\centerline{\includegraphics[width=3.5in]{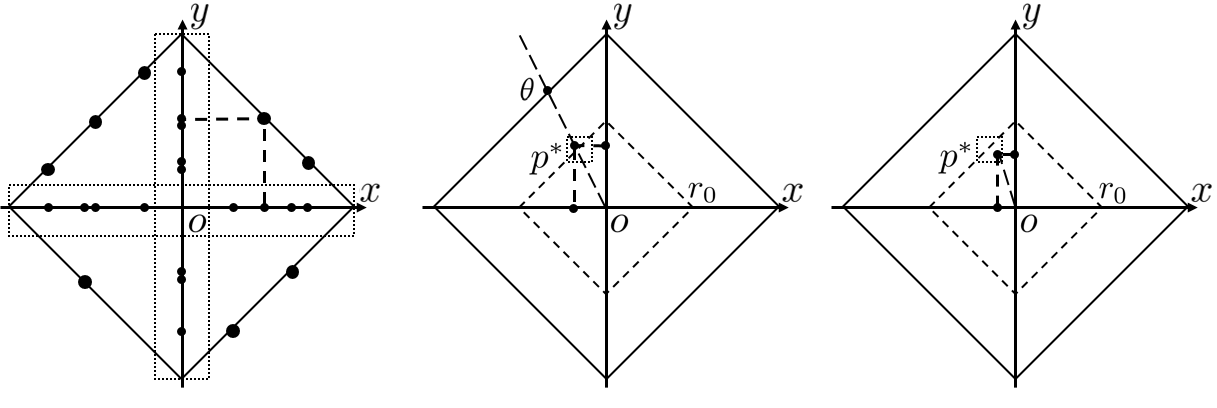}}
\caption{The $1$-norm realization on $\mathbb R^2$.}
\label{fig:ftcc1}
\end{figure}

In considering the realization, to avoid trigonometric and square root operations, we can use the $1$-norm circle shown in the rightmost of Fig.~\ref{fig:ftcc0} on the plane $P$ rather than the $2$-norm circle to represent the unified circle $C$ with the same circumference $1$.
Namely, the diamond $D=aba'b'$ is homeomorphic to $C$.
As $|C|=1$, we have $\mathring{d}(a,b)=\mathring{d}(b,a')=\mathring{d}(a',b')=\mathring{d}(b',a)=1/4$ on $D$.
During the first step, points of $S_q$ are located on $D$ with their directed distances to point $a$ being maintained ($a$ is also the point $c_\mathtt{0}=0$ on $C$), as is shown in the leftmost of Fig.~\ref{fig:ftcc1}.
Then, in running the $\mathtt{Resync}$ algorithm shown in Fig.~\ref{fig:algoCFResync}, the projections of $S_q$ to the $ox$ and $oy$ axes are iteratively handled by the FTA function during the $k_1$th ($1\leqslant k_1\leqslant K_1$) steps to generate the anchor point $p^*=(x^*,y^*)$ at the $K_1$th step, as is shown in Fig.~\ref{fig:ftcc1}.

In simplifying the computation, denoting the $1$-norm distance between points $u=(x_1,y_1)$ and $v=(x_2,y_2)$ on $P$ as $|uv|_1=|x_1-x_2|+|y_1-y_2|$, the intersection point $c_\mathtt{1}$ can be represented as $c_\mathtt{1}=(\frac{|oc_\mathtt{1}|_1}{|op^*|_1}x^*,\frac{|oc_\mathtt{1}|_1}{|op^*|_1}y^*)$ with $|oc_\mathtt{1}|_1=|oa|_1$.
As $c_\mathtt{1}$ is also a point on $D$, it corresponds to a unique counterclockwise length $c_\mathtt{1}=(i-1+(((i-1)\%2)|x^*|+(i\%2)|y^*|)/|op^*|_1)/4$ from $c_\mathtt{0}$ to $c_\mathtt{1}$ along $D$ when $c_\mathtt{1}$ is at the $i$th ($1\leqslant i \leqslant 4$) quadrant of the $oxy$ coordinate system on $P$.
Then, $c_\mathtt{w}$ would be determined by running the $K_2$-step routine $\mathtt{X}$ with $r_q$ being the input.
Concretely, when $\mathtt{X}$ is realized with $\mathtt{A}$, the $K_2$-step routine $\mathtt{A}$ would run in $q$ with the boolean input $(r_q> r_0)$ and the boolean output $c_\mathtt{w}\in\{0,1\}$.
Otherwise, when $\mathtt{X}$ is realized with exchanging the random boolean numbers, $c_\mathtt{w}$ would be computed with (\ref{eq_c_w}) and (\ref{eq_b_w}).

To differentiate the realizations, we use $\mathtt{ResyncA}$ and $\mathtt{ResyncR}$ to represent the $\mathtt{Resync}$ algorithms realized with and without the $\mathtt{A}$ protocol, respectively.
In all cases, $c^*$ would be computed with (\ref{eq_creference}).
It should be noted that, as we take the $1$-norm distance on $P$, the circumference of $D$ would be measured as $8|oa|_1$ on $P$.
Meanwhile, the $1$-norm distances are unified to avoid unnecessary division operations.
Also, the computation would be performed on $\theta_{r,q}$ rather than $\theta_{r,q}/T$.
Thus, the algorithm needs only to manipulate integers, which is preferred in most digital systems.

\alglanguage{pseudocode}
\algrenewcommand{\algorithmiccomment}[1]{\hskip1em//#1}
\begin{figure}[htbp]
\centering
\begin{algorithmic}[1]
\Statex \underline{the first stage:}
    \State  send $\theta_{q}$ and receive $\theta_{r,q}$ from $r\in N$;
    \State  $S_q:=\{\theta_{r,q}/ T\mid r\in N\}$;
            $s_q:=\theta_{q}/ T$;
    \State  $S_{q.x}:=\{1/4-\mathring{d}(c,0) \mid c\in S_q\}$;
    \State  $S_{q.y}:=\{1/4-\mathring{d}(c,1/4) \mid c\in S_q\}$;
    \State  $s_{q.x}:=\mathtt{FTA}(S_{q.x})$;
            $s_{q.y}:=\mathtt{FTA}(S_{q.y})$;
    \For{$k_1=2:K_1$}
        \State  send $(s_{q.x},s_{q.y})$ and receive $(s_{r.x},s_{r.y})$ from $r\in N$;
        \State  $S_{q.x}:=\{s_{r.x}\mid r\in N\}$;
                $S_{q.y}:=\{s_{r.y}\mid r\in N\}$;
        \State  $s_{q.x}:=\mathtt{FTA}(S_{q.x})$;
                $s_{q.y}:=\mathtt{FTA}(S_{q.y})$;
    \EndFor
\Statex \underline{the second stage:}
    \State  $r_q:=|s_{q.x}|+|s_{q.y}|$;
            $c_\mathtt{1}:=0$;
            $c_\mathtt{w}:=0$;
    \If{$r_q> 0$}
        \State $c_\mathtt{1}:=((s_{q.y}\geqslant 0)?(1-s_{q.x}/r_q):(3+s_{q.x}/r_q))/4$;
    \EndIf
    \State  $c_\mathtt{w}=\mathtt{X}(r_q)$;
    \State  $c^*:=c_\mathtt{w} c_\mathtt{1}$;
    \State \Return $\Delta \theta_{q}:= T((c^*\ominus s_q<1/2)?c^*\ominus s_q:-(s_q\ominus c^*))$;
\end{algorithmic}
\caption{The $\mathtt{Resync}$ algorithm for node $q$.}\label{fig:algoCFResync}
\end{figure}

Alternatively, the $\infty$-norm circle on the plane $P$ would also do with some straightforward intuition.
Here we leave this to the interested readers.

\section{Analysis}
\label{sec:Ana}

For simplicity, we assume that the $\mathtt{FTA}$ function is the synchronous approximate function provided in \cite{RN4093}.
Meanwhile, the $\mathtt{A}$ routine is the $P_r$ protocol provided in \cite{FM1997}.
With Lemma~8 of \cite{RN4093}, the \emph{Convergence} property of the $\mathtt{FTA}$ function is directly employed with $\mathtt{c}=\lfloor (n-2f-1)/f \rfloor+1$.
As $n>3f$, we have $\mathtt{c}\geqslant 2$.
And with Claim T4-4 of \cite{FM1997} (see also Lemma~30 of \cite{Lenzen2019AlmostConsensus}), the \emph{Deterministic Validity} (\emph{Validity} in short) property and the \emph{Probabilistic Consistency} (or saying the \emph{Probabilistic Agreement}) property \citep{Lenzen2019AlmostConsensus} of the $\mathtt{A}$ routine is also directly employed with the probability $0.35$.

For analysis, points of all nonfaulty nodes in $S_q$ are denoted as $S_q^{Q}=\{c_{r,q}\mid r\in Q\}$.
A local variable $x$ in the node $q\in Q$ is denoted as $x^{(q)}$.
For example, the anchor point, ashore point, and reference point of node $q$ are respectively denoted as $p^{*(q)}$, $c_\mathtt{1}^{(q)}$, and $c^{*(q)}$.
Besides $\epsilon_1$ and $\epsilon_2$, other related parameters are collected here as $\epsilon_0=2\rho(T_0+\Delta_0)/ T$, $\epsilon_1'=\epsilon_1+\epsilon_0+(1+\rho)\delta_s$, $\varepsilon_0=4\epsilon_1'|oa|_1$, $\varepsilon_1=\mathtt{c}^{-K_1}\varepsilon_0$, $\varepsilon_2=2\mathtt{c}^{-K_1}|oa|_1$.

\subsection{Synchronized states}
Firstly, we show that $\mathtt{ResyncA}$ and $\mathtt{ResyncR}$ can satisfy the \emph{convergency} and the \emph{continuity} properties required in (\ref{eq_Convergency}) and (\ref{eq_Continuity}) deterministically when $\mathcal{S}$ is synchronized in the presence of $f<n/3$ Byzantine nodes.
\begin{lemma}
\label{lemma_continuity}
If $|\iota_Q(t_{last}^{(k-1)+})|\leqslant \epsilon_1$, $\mathtt{c}^{-K_1}\frac{\epsilon_1'}{(1-4\epsilon_1')}+(1+\rho)\delta_s\leqslant\epsilon_1<1/4-\epsilon_0-(1+\rho)\delta_s$ and $\varepsilon_1<r_0<r_0'< |oa|_1-\varepsilon_0$, then $|\iota_Q(t_{last}^{(k)+})|\leqslant \epsilon_1$ and $\iota_Q(t_{last}^{(k)+})\subseteq \iota_Q(t_{first}^{(k)-})+\kappa(a_k,\rho\delta_s+\epsilon_1')$.
\end{lemma}
\begin{IEEEproof}
As $|\iota_Q(t_{last}^{(k-1)+})|\leqslant \epsilon_1$, the points in $S_q^{Q}$ of every node $q\in Q$ in the $k$th r-round are within an $\varepsilon_0$-square $U_0$ centered on some point of $D$.
Thus, the corresponding projections $S_{q.x}^{1}$ and $S_{q.y}^{1}$ of the nodes in $Q$ can be covered by a $\varepsilon_0$-length interval on the $ox$ and $oy$ axes respectively.
Then, by iteratively applying the \emph{Convergence} property of the $\mathtt{FTA}$ function, the anchor points $p^*$ of all nodes in $Q$ would be covered by an $\varepsilon_1$-square $U\subseteq U_0$.
As $\epsilon_1<1/4-\epsilon_0-(1+\rho)\delta_s$, we have $\forall p\in U: |op|_1\geqslant|oa|_1-\varepsilon_0$.
As $r_0< r_0'< |oa|_1-\varepsilon_0$, every $q\in Q$ would get $(r_q>r_0'>r_0)$ being true.
Thus, by applying the \emph{Deterministic Validity} property of the $\mathtt{A}$ routine, the reference point of every node $q\in Q$ would be set as $c_\mathtt{1}^{(q)}$ with $\mathtt{ResyncA}$.
Meanwhile, with (\ref{eq_c_w}), $\mathtt{ResyncR}$ would act as the same of $\mathtt{ResyncA}$.
As $r_0>\varepsilon_1$, for any two nodes $q_1,q_2\in Q$ with $c_\mathtt{1}^{(q_1)}$ and $c_\mathtt{1}^{(q_2)}$ being on the same quadrant of $P$, we have $|c_\mathtt{1}^{(q_1)}c_\mathtt{1}^{(q_2)}|_1\leqslant\frac{|oa|_1}{|oa|_1-\varepsilon_0}|p^{*(q_1)}p^{*(q_2)}|_1<2\varepsilon_1/(1-4\epsilon_1')$ on $P$.
So in all cases $\mathring{d}(c_\mathtt{1}^{(q_1)},c_\mathtt{1}^{(q_2)})=\frac{1}{8|oa|_1}|c_\mathtt{1}^{(q_1)}c_\mathtt{1}^{(q_2)}|_1< \frac{2\varepsilon_1/(1-4\epsilon_1')}{8|oa|_1}=\mathtt{c}^{-K_1}\frac{\epsilon_1'}{(1-4\epsilon_1')}$ holds.
Thus $|\iota_Q(t_{last}^{(k)+})|\leqslant \epsilon_1$ holds.
As $U\subseteq U_0$, we have $\forall q\in Q:\Delta \theta_{q}^{k}\leqslant \epsilon_1' T$.
Thus we have $\epsilon_2\leqslant\epsilon_1'$.
\end{IEEEproof}

\subsection{Arbitrary states}
Then, we show that the \emph{convergency} property required in (\ref{eq_Convergency}) can be satisfied with some fixed probabilities with arbitrary system states.
Firstly, we show $\mathtt{ResyncA}$ can satisfy the \emph{convergency} property with a probability no less than $0.35$ in the presence of $f<n/3$ Byzantine nodes.
\begin{lemma}
\label{lemma_convergencyA}
With $\mathtt{ResyncA}$, $\epsilon_1\geqslant 1/(4r_0/\varepsilon_2-4)+\epsilon_0$, $r_0>\varepsilon_2$, and the other parameters being set as in Lemma~\ref{lemma_continuity}, for every $k\geqslant 1$ there is at least a probability $0.35$ that $|\iota_Q(t_{last}^{(k)+})|\leqslant \epsilon_1$ holds.
\end{lemma}
\begin{IEEEproof}
With the proof of Lemma~\ref{lemma_continuity}, the anchor points of all nodes in $Q$ would be covered by an $\varepsilon_2$-square $U$ on $P$.
Denoting the center point of $U$ as $u$, if $|ou|_1<r_0-\varepsilon_2$,  $\forall q\in Q:|op^{*(q)}|_1<r_0$ holds and thus $\forall q\in Q:r_q<r_0$ holds.
In this case, by applying the \emph{Deterministic Validity} property of the $\mathtt{A}$ routine, the reference point of every node $q\in Q$ would be set as $c^{*(q)}=c_\mathtt{0}$.
Otherwise, if $|ou|_1\geqslant r_0-\varepsilon_2$, as $r_0 >\varepsilon_2$, we have $\forall q_1,q_2\in Q:\mathring{d}(c_\mathtt{1}^{(q_1)},c_\mathtt{1}^{(q_2)})=\frac{1}{8|oa|_1}|c_\mathtt{1}^{(q_1)}c_\mathtt{1}^{(q_2)}|_1\leqslant\frac{1}{8(r_0-\varepsilon_2)}|p^{*(q_1)}p^{*(q_2)}|_1<\frac{\varepsilon_2}{4(r_0-\varepsilon_2)}$.
So if $\mathtt{A}$ outputs $1$ in all nodes of $Q$, $\forall q\in Q:c^{*(q)}=c_\mathtt{1}^{(q)}$ holds and thus $\forall q_1,q_2\in Q:\mathring{d}(c^{*(q_1)},c^{*(q_2)})\leqslant\frac{1}{4(r_0/\varepsilon_2-1)}$.
Otherwise, if $\mathtt{A}$ outputs $0$ in all nodes of $Q$, we have $\forall q\in Q:c^{*(q)}=c_\mathtt{0}$.
So by applying the \emph{Probabilistic Consistency} property of the $\mathtt{A}$ routine, $\forall q_1,q_2\in Q:\mathring{d}(c^{*(q_1)},c^{*(q_2)})\leqslant \frac{1}{4(r_0/\varepsilon_2-1)}$ holds with at least a probability $0.35$.
So, as the accumulated clock drifts of every node in $Q$ would not exceed $\epsilon_0/2$, $|\iota_Q(t_{last}^{(k)+})|\leqslant \epsilon_1$ holds with at least a probability of $0.35$.
\end{IEEEproof}

Now we show $\mathtt{ResyncR}$ can satisfy the \emph{convergency} property with a fixed positive probability in the presence of $f=O(\sqrt{n})$ Byzantine nodes.
For this, we first restate the classical result of random walks (the basic and some extended results can all be found in \citep{Rayleigh1880}) for our case.
\begin{lemma}\citep{Rayleigh1880}
\label{lemma_random_walk_r}
If $|Q|$ is enough large, $\tau\in[0,1/\sqrt{2}]$, and $b_\mathtt{w}^{(q)}$ being independent of $b_\mathtt{w}^{(q')}$ for all $q,q'\in Q$, then there is approximately a probability $1-\frac{2}{\sqrt{\pi}}\int_{0}^{\tau}e^{-\tau^2}\mathrm {d}t$ that $|\sum_{i\in Q}b_\mathtt{w}^{(i)}|\geqslant 2\tau\sqrt{|Q|/2}$.
\end{lemma}
\begin{IEEEproof}
See \cite{Rayleigh1880}.
\end{IEEEproof}

For example, when $|Q|$ is large, as is shown in \cite{Rayleigh1880}, the probability of $|\sum_{i\in Q}b_\mathtt{w}^{(i)}|\geqslant \sqrt{|Q|/2}$ can be estimated as $1-\frac{2}{\sqrt{\pi}}\int_{0}^{1/2}e^{-\tau^2}\mathrm {d}t\approx 0.4795$.
With ignoring the estimation errors, we provide the following approximate observation.

\begin{lemma}
\label{lemma_convergencyR}
With $\mathtt{ResyncR}$, if $n$ is sufficiently large, $f< \sqrt{(n-f)/2}$, $r_0'-r_0>\varepsilon_1$, and the other parameters being set as in Lemma~\ref{lemma_convergencyA}, then for every $k\geqslant 1$ there is at least a probability $0.2397$ that $|\iota_Q(t_{last}^{(k)+})|\leqslant \epsilon_1$ holds.
\end{lemma}
\begin{IEEEproof}
As $|Q|\geqslant n-f$, with a probability $0.4795$ that $|\sum_{i\in Q}b_\mathtt{w}^{(i)}|\geqslant \sqrt{(n-f)/2})>f$.
In this case, with (\ref{eq_c_w}) and (\ref{eq_b_w}), we would have $\forall q_1,q_2\in Q:c_\mathtt{w}^{(q_1)}=c_\mathtt{w}^{(q_2)}$.
Meanwhile, as every $b_\mathtt{w}^{(q)}$ is the result of an unbiased coin, for every $b\in\{0,1\}$, the probability of $\forall q\in Q:c_\mathtt{w}^{(q)}=b$ is no less than $0.4795/2>0.2397$.
Thus, with $r_0'-r_0>\varepsilon_1$ and $\forall q_1,q_2\in Q: |r_{q_1}-r_{q_2}|\leqslant \varepsilon_1$, if there exists $q\in Q$ satisfying $r_{q}\geqslant r_0'$, the probability of $\forall q\in Q:c_\mathtt{w}^{(q)}=1$ is no less than $0.2397$.
Also, if there exists $q\in Q$ satisfying $r_{q}\leqslant r_0$, the probability of $\forall q\in Q:c_\mathtt{w}^{(q)}=0$ is no less than $0.2397$.
Otherwise, when $\forall q\in Q:r_{q}\in (r_0,r_0')$, the probability of $\forall q\in Q:c_\mathtt{w}^{(q)}$ is no less than $0.4795$.
So, in all cases, as the accumulated clock drifts of every node in $Q$ would not exceed $\epsilon_0/2$, $|\iota_Q(t_{last}^{(k)+})|\leqslant \epsilon_1$ holds with at least a probability $0.2397$.
\end{IEEEproof}

For small $n$, the result of Lemma~\ref{lemma_convergencyR} still holds similarly or even better.
For example, when $n=5$ and $f=1$, as the probability of $|\sum_{i\in Q}b_\mathtt{w}^{(i)}|> f$ is no less than $0.625$, $|\iota_Q(t_{last}^{(k)+})|\leqslant \epsilon_1$ would hold with a probability no less than $0.312$.

\subsection{Solving the parameters}
To satisfy the restrictions of the parameters, we can set
\begin{eqnarray}
\label{eq_set1}
\frac{r_0}{|oa|_1}\in(2\mathtt{c}^{-K_1}(\frac{1}{4(\epsilon_1-\epsilon_0)}+1),1-4\epsilon_1')\\
\label{eq_set2}
\epsilon_1\in [\mathtt{c}^{-K_1}\frac{\epsilon_1'}{1-4\epsilon_1'}+(1+\rho)\delta_s,\frac{1}{4}-\epsilon_0-(1+\rho)\delta_s)
\end{eqnarray}

As $\mathtt{c}\geqslant 2$, (\ref{eq_set1}) and (\ref{eq_set2}) can be easily satisfied with a sufficiently large $K_1$ and sufficiently small $\rho$ and $\delta_s$.
For example, if $\max\{2\epsilon_0,2(1+\rho)\delta_s,\epsilon_1\}\leqslant 1/32$, we would have $\epsilon_1'<1/16$ and thus $\epsilon_1\in [\mathtt{c}^{-K_1}/12+(1+\rho)\delta_s,1/32]$ is required in satisfying (\ref{eq_set2}).
So by setting $K_1\geqslant \lceil \log_\mathtt{1/c}12(1/32-(1+\rho)\delta_s) \rceil$, we can always set $\epsilon_1=\mathtt{c}^{-K_1}/12+(1+\rho)\delta_s\leqslant 1/32$.
Then, by setting $K_1\geqslant \lceil \log_\mathtt{c}(2/3)(1/(\epsilon_1-\epsilon_0)+4) \rceil+1$, we can set $\frac{r_0}{|oa|_1}\in(\mathtt{c}^{-K_1}(1/(\epsilon_1-\epsilon_0)+4)/2,3/4)$ in satisfying (\ref{eq_set1}).
Thus, $r_0$ and $r_0'$ can be easily solved in this interval.

\subsection{Main results}
Now we show that the desired pulsing system $\mathcal{S}_1$ can be established in several r-rounds with a high probability.
For $\mathtt{ResyncA}$, we assume $f<n/3$ and $\eta=0.35$.
For $\mathtt{ResyncR}$, we assume $f<\sqrt{(n-f)/2}$ and $\eta=0.2397$.
As is shown in the preceding examples, the concrete Byzantine resilience of $\mathtt{ResyncR}$ can be better than that of the general case.

\begin{theorem}
\label{theorem_randomized}
With at least a probability $1-(1-\eta)^k$ that $\mathcal{S}$ would be a $(\Pi, T, \Delta)$-pulsing system since $k(T_0+\Delta_0)$ with $\Pi= (\epsilon_0+\epsilon_1)T$ and $\Delta= \lceil (T+\Delta)/(T_0-\Delta_0)\rceil\epsilon_1'T+\rho (T+\Delta)$.
\end{theorem}
\begin{IEEEproof}
Firstly, as $\mathcal{S}_0$ is a $(\Pi_0, T_0, \Delta_0)$-pulsing system, every r-round of $\mathcal{S}$ would not be overlapped with any other r-round of $\mathcal{S}$ and at least $k$ r-rounds would be executed before $k(T_0+\Delta_0)$.
During the $i$th r-round, by applying Lemma~\ref{lemma_convergencyA} and Lemma~\ref{lemma_convergencyR} , with at least a probability $\eta$ that $|\iota_Q(t_{last}^{(i)+})|\leqslant \epsilon_1$ holds.
And once $|\iota_Q(t_{last}^{(i)+})|\leqslant \epsilon_1$ holds, by applying Lemma~\ref{lemma_continuity}, $|\iota_Q(t_{last}^{(i+1)+})|\leqslant \epsilon_1$ and $\iota_Q(t_{last}^{(i+1)+})\subseteq \iota_Q(t_{first}^{(i+1)-})+\kappa(a_{i+1},\rho\delta_s+\epsilon_1')$ hold.
In this case, as the scheduled pulsing tick $\tau_{sch}^{(q)}$ can only be adjusted (not reset) by the end of every r-round in every $q\in Q$, the \emph{Precision} property required in (\ref{eq_sync_precision}) can be satisfied with $\Pi$.
And as there are at most $\lceil (T+\Delta)/(T_0-\Delta_0)\rceil$ r-rounds between any two adjacent pulses in every node $q\in Q$, the \emph{Accuracy} property required in (\ref{eq_sync_accuracy}) can be satisfied with $\Delta$.
So with at least a probability $1-(1-\eta)^k$ that $\mathcal{S}$ would be a $(\Pi, T, \Delta)$-pulsing system since $k(T_0+\Delta_0)$.
\end{IEEEproof}

So, by executing the $\mathtt{Resync}$ routine $k$ times, SSBPR can be reached with a probability no less than $1-(1-\eta)^k$.
Meanwhile, as the $\mathtt{Resync}$ routine can also maintain the synchronized state of the system, SSBPS is also reached at the same time with the \emph{igc} pulses.
So with the \emph{igc} pulses, both \emph{ECT} SSBPR and \emph{ECT} SSBPS can be reached with an expected constant time no more than $\Gamma_0/\eta<5K_s \Phi_{s}^{+}$ with $\Phi_{s}^{+}$ being approximately $2\Pi_0+d$ and $K_s$ being independent of $f$ and $n$.
Concretely, with $\mathtt{ResyncA}$, \emph{ECTOR} SSBPR can be reached at the expense of high message complexity.
With $\mathtt{ResyncR}$, SSBPR can be reached in fewer rounds ($K_1+1$ rounds) with very low message complexity (like that of approximate agreement) in the presence of $O(n)$ Byzantine nodes.
Moreover, as is shown in the preceding examples, the SSBPR solution provided with employing $\mathtt{ResyncR}$ can make a better tradeoff between Byzantine tolerance and algorithm efficiency even when $n$ is small.

\section{Conclusion}
\label{sec:Conclusion}
This paper explored the ECT SSBPR problem with preferring geometric operations, approximate agreements, random walks, and other constant-time operations.
Firstly, by relaxing the common pulses as the \emph{igc} pulses, the classical digital SSBCS problem is extended to the real-time SSBPR problem.
Then, the ECT SSBPR problem is investigated with some geometric intuitions.
Concretely, two expected-constant-time SSBPR solutions are provided with integrating geometric embedding, approximate agreement, probabilistic consensus, and random walks.
With the SSBPR solutions and the \emph{igc} pulses, \emph{ECT} SSBPR and \emph{ECT} SSBPS are reached without taking large time coefficients.
Providing that the \emph{igc} pulses can be efficiently generated with some higher assumption coverage than that of the regularly generated pulses, the provided SSBPR solution can efficiently convert \emph{igc} pulses to practical \emph{ECT} SSBPS.
Also, the questions posed in \cite{Ben2008Fast} and \cite{Lenzen2019AlmostConsensus} are now narrowed to show if there is \emph{ECT} solutions that can generate the \emph{igc} pulses in the bounded-delay model.

Despite the merits, several obstacles still exist in preventing the SSBPR solutions from being widely applied.
Firstly, as the message complexity of the constant-round probabilistic consensus is high, the provided \emph{ECTOR} solution can only be applied in high-bandwidth communication networks.
In overcoming this, the probabilistic consensus routine can be replaced as some deterministic consensus routine, but the stabilization time would be linear to $f$.
Secondly, the provided SSBPS solutions still rely on the \emph{igc} pulses.
To provide the \emph{igc} pulses in safety-critical systems, the designer should show sufficiently high assumption coverage of the underlying pulsing systems.

\bibliographystyle{IEEEtran}
\bibliography{IEEEabrv,ECTSSBPR}

\end{document}